\documentclass[letterpaper]{article} 
\usepackage{aaai2026}  
\usepackage{times}  
\usepackage{helvet}  
\usepackage{courier}  
\usepackage[hyphens]{url}  
\usepackage{graphicx} 
\usepackage{natbib}  
\usepackage{caption} 
\usepackage{algorithm}
\usepackage{algorithmic}
\usepackage{booktabs}
\usepackage{makecell}
\usepackage{amsmath}
\usepackage{multirow}
\usepackage{subcaption}
\usepackage{spverbatim}

\usepackage{xcolor}
\newcommand{\answerYes}[1]{\textcolor{blue}{#1}} 
\newcommand{\answerNo}[1]{\textcolor{teal}{#1}} 
\newcommand{\answerNA}[1]{\textcolor{gray}{#1}} 
 
\usepackage{newfloat}
\usepackage{listings}
\DeclareCaptionStyle{ruled}{labelfont=normalfont,labelsep=colon,strut=off} 
\floatstyle{ruled}
\newfloat{listing}{tb}{lst}{}
\floatname{listing}{Listing}
\title{Asymmetric Dynamics of Partisan Warriors in YouTube Comments}
\author{
    Keyeun Lee\textsuperscript{\rm 1},
    Sang Jung Kim\textsuperscript{\rm 2}
}
\affiliations{
    \textsuperscript{\rm 1}Seoul National University\\
    \textsuperscript{\rm 2}University of Iowa

    kieunp@snu.ac.kr, sangjung-kim@uiowa.edu
}

\usepackage{bibentry}

\begin{document}

\maketitle

\begin{abstract}
Cross-cutting commenting on social media is often imagined as a path to deliberation, yet exposure to opposing views frequently fuels hostility. To explain this dynamic, we introduce the concept of partisan warriors--commenters who cross ideological lines primarily to launch uncivil attacks against out-partisans. We analyze a large corpus of YouTube comments (N= 1,854,320) surrounding the 2024 U.S. second presidential debate. After filtering for toxicity and active participation, we use large language models to identify attack targets and operationalize partisan warrior behavior. Our analysis highlights four dynamics. First, cross-cutting commenters do not exhibit greater civility than those who remain within their ideological camps (RQ1). Second, audience reactions diverge by ideology: conservative audiences tended to reward hostile attacks on out-group leaders, whereas liberal audiences offered no comparable incentives and at times penalized such attacks (RQ2). Third, partisan warriors are notably more prevalent in conservative-leaning channels than in liberal ones; commenters restricted to conservative spaces were substantially more likely to engage in partisan warrior behavior compared to their liberal-only counterparts (RQ3). Finally, regarding environmental triggers, robustness checks suggest that this participation is an ecological phenomenon driven largely by channel-level heterogeneity rather than transient responses to individual video titles (RQ4). By shifting attention from the prevalence of incivility to its targets, rewards, and structural drivers, this study advances understanding of how partisan hostility is enacted and sustained in online spaces.
\end{abstract}


\section{Introduction}
Social media platforms are widely criticized for reinforcing echo chambers and filter bubbles, limiting exposure to opposing perspectives along partisan lines \cite{bakshy_exposure_2015, flaxman_filter_2016, sunstein_republic_2018}. Against this backdrop, cross-cutting commenters--those who engage across partisan divides--are often seen as embodying a more democratic ideal: the promise of deliberation, tolerance, and dialogue \cite{garrett_echo_2009, mutz_cross-cutting_2002, wojcieszak_online_2009}. Yet in practice, cross-commenting often manifests as antagonism and incivility toward out-partisans, raising questions about whether such encounters bridge divides or deepen partisan hostility. Rather than bridging divides, cross-cutting commenters may function as performers of partisan hostility \cite{cetinkaya_cross-partisan_2025, wu_cross-partisan_2021, xia_integrated_2025}.

While the literature on incivility has primarily focused on the degree of hostility in individual posts or comments \cite{coe_online_2014, rossini_beyond_2022, theocharis_bad_2016, yudhoatmojo_data-driven_2024}, a fuller understanding requires examining its targets \cite{matsui_hater_2024}. Incivility toward political outgroups reflects proactive denigration, making it important to distinguish between commenters who remain within ideologically aligned spaces and those who deliberately cross partisan lines to criticize opponents \cite{kim_online_2018}.

Despite growing interest in political incivility and cross-cutting contact, a key gap remains: we lack a systematic way to identify commenters who cross ideological boundaries primarily to attack out-partisans. Prior work has examined the prevalence of incivility and the structure of cross-cutting interactions \cite{wu_cross-partisan_2021, cetinkaya_cross-partisan_2025}, but it has paid little attention to the targets of incivility or to the audience responses that may incentivize and sustain such behavior.

We address this gap by introducing the concept of \textit{partisan warriors}--commenters who leave uncivil remarks directed at out-partisans in opposing channels. This conceptualization matters because crossing into hostile territory is atypical in selective-exposure environments and signals a confrontational form of partisan expression \cite{garrett_echo_2009, rossini_beyond_2022, stroud_polarization_2010}. Studying partisan warriors allows us to move beyond measuring the amount of incivility to examining who is targeted and how partisan identity is performed in cross-cutting contexts. In this paper, we compare the prevalence of partisan warriors among commenters active in conservative channels, cross-commenters, and commenters active in liberal channels on YouTube.

Beyond commenters themselves, we also consider the broader platform dynamics that shape partisan conflict. The toxicity of videos may encourage or deter partisan warriors, while the ways in which audiences reward or sanction comments reveal whether partisan hostility is amplified or constrained \cite{theocharis_bad_2016, gao_crisis_2024}. Together, these dynamics clarify not only the behavior of commenters but also the broader processes through which platforms sustain or exacerbate partisan conflict.

Building on this motivation, our study seeks to unpack the layered relationships that structure partisan discourse: the channels, the commenters who engage with them, the videos channels host, the comments those videos generate, and the audiences who react. This multi-level perspective allows us to move beyond isolated analyses of incivility and instead trace how partisan hostility is produced, expressed, and rewarded across the platform. Accordingly, we focus on four research questions:

\begin{itemize}
  \item[] \textbf{RQ1} Is there a difference in the level of incivility between cross-cutting commenters and commenters who remain within ideologically aligned spaces? 
    \item[] \textbf{RQ2} How do audiences in conservative and liberal channels reward or sanction uncivil comments directed at out-partisans?  
  \item[] \textbf{RQ3} How does the prevalence of partisan warriors differ across conservative-channel-only commenters, cross-commenters, and liberal-channel-only commenters?  
  \item[] \textbf{RQ4} How does the toxicity of videos influence the participation of partisan warriors in comment sections?  
\end{itemize}

\section{Related Work}
\subsection{Cross-cutting Exposure and Incivility}
Cross-cutting exposure--the encounter with opposing political views--has long been theorized as essential for democratic deliberation \cite{mutz_cross-cutting_2002}. Online platforms make such contact observable at scale, as users encounter dissonant perspectives in comment threads and social media feeds \cite{bakshy_exposure_2015}. Yet its democratic value remains contested. Some studies suggest that exposure fosters tolerance and reduces extremity \cite{mutz_cross-cutting_2002}, while others find it provokes defensiveness, selective avoidance, or withdrawal \cite{stroud_polarization_2010}. More recent work shows that contact with opposing views can heighten political polarization \cite{bail_exposure_2018}.

These tensions are especially visible in comment sections, where cross-cutting engagement often devolves into antagonism and incivility rather than reasoned dialogue \cite{masullo_toxic_2023}. In some cases, both deliberation and incivility rise together, suggesting that cross-cutting encounters are double-edged \cite{rossini_characterizing_2021}. Recent work on Twitter similarly shows that cross-partisan interactions are relatively rare and, when they occur, tend to be negative rather than constructive \cite{cetinkaya_cross-partisan_2025}.

Prior work has documented strong connections between toxic or hostile language in political discourse and affective polarization on social media \cite{lerman2024affective}. Interactions between users with opposing ideological affiliations are more likely to exhibit negativity and toxicity, consistent with affective polarization in online exchanges \cite{lerman2024affective, simchon2022troll}. Recent computational research further emphasizes the role of toxic and uncivil speech in shaping polarized dynamics in cross-partisan interactions. Large-scale audits of social media ranking systems show that engagement-based algorithms can amplify hostile and divisive content, motivating interest in alternative ranking approaches that prioritize prosocial or constructive interactions \cite{milli2024engagement}. Related efforts, including the Twitter Prosocial Ranking Challenges, examine whether re-ranking content based on prosocial attributes can reduce exposure to toxic speech and improve the quality of cross-partisan engagement \cite{prosocial_ranking_challenge_2024, saltz2024reranking}. These studies highlight how platform design and content ranking shape whether exposure to opposing viewpoints leads to deliberation or incivility.

Beyond explicitly political contexts, prior research shows that the discourse of highly engaged partisans is often characterized by heightened toxicity even when political topics are not salient \cite{mamakos2023social}. Such findings further motivate a behavioral focus on incivility, as antagonistic rhetoric may emerge independently of issue content or informational disagreement. Uncivil cross-partisan engagement can be understood as an expression of partisan identity rather than a response to specific political claims.

Building on this scholarship, our work directly compares the incivility of cross-cutting commenters with that of commenters operating within partisan-aligned channels. By focusing on observable rhetorical behavior, we provide systematic evidence on whether cross-cutting engagement is uniquely associated with heightened incivility.

\subsection{Audience Reactions and the Rewarding of Incivility}
Comments from both cross-cutting commenters and those in partisan-aligned channels are not consumed in isolation; they are evaluated by audiences whose responses--likes, upvotes, and replies--signal which voices are legitimized within partisan communities. Prior work shows that incivility can be rewarded as authentic or entertaining \cite{rossini_beyond_2022}, but it can also be sanctioned as norm-violating \cite{theocharis_dynamics_2020}. Platform affordances further shape these dynamics: algorithmic amplification often privileges ``engaging'' content regardless of its tone, thereby elevating hostile participation \cite{ribeiro_auditing_2020}.

Recent research adds to this picture by modeling how different categories of incivility affect conversational outcomes and user engagement \cite{gao_crisis_2024}, suggesting that audiences play an active role in sustaining or constraining hostile discourse. Our contribution lies in examining whether partisan hostility is rewarded differently when comments are directed at political ingroups or outgroups, using LLMs to attribute incivility to its partisan targets and to connect these patterns with audience responses.

\subsection{Targets of Incivility and Partisan Warriors}
Research on political incivility has largely examined its forms and prevalence. Early typologies distinguished insults, name-calling, and pejorative speech \cite{papacharissi_democracy_2004, coe_online_2014}, while later work emphasized its multidimensional character \cite{stryker_what_2016, pendzel_closer_2024}. Computational social science has extended these approaches by operationalizing toxicity for large-scale detection, often through machine learning models such as the Perspective API, which assigns toxicity scores based on linguistic cues \cite{almerekhi_detecting_2019, gao_crisis_2024}.

Far fewer studies investigate who is targeted by incivility. Beyond identification, incivility also serves functional roles when directed at specific groups: it can polarize by delegitimizing opponents \cite{gervais_incivility_2015, theocharis_dynamics_2020}, but also mobilize by energizing supporters and signaling authenticity \cite{rossini_beyond_2022}. To advance this line of work, we leverage large language models (LLMs), which enable scalable annotation of incivility across millions of comments while preserving contextual nuance. Unlike dictionary approaches or toxicity classifiers, LLMs can incorporate surrounding conversational context and be prompted to recognize who is targeted, allowing us to distinguish hostility toward out-partisans from generic uncivil speech. Recent work has begun to demonstrate the feasibility of using LLMs for nuanced annotation tasks in social media research \cite{farjam_practical_2025}, underscoring their potential for advancing incivility studies.

We introduce the concept of \textbf{Partisan Warriors (PWs)}: commenters who leave uncivil remarks directed at out-partisans in opposing channels. Conceptually, PWs are closely related to political trolls, as both engage in antisocial behavior toward political opponents and contribute to polarized political discussions on social media \cite{fichman2023trolling}. Prior work on political trolling documents practices such as provocation and the disruption of political discourse through uncivil interactions with opposing partisans \cite{fichman2023trolling, simchon2022troll}. At the same time, political trolling is often operationalized to include the dissemination of misinformation or disinformation, as well as strategic or deceptive behavior intended to manipulate political conversations \cite{fichman2023trolling, blinder_ware_2024_sweden_troll_farm}. In contrast, our operationalization of partisan warriors does not incorporate the spread of mis- or disinformation. Instead, we focus on observable rhetorical behavior on social media, irrespective of informational accuracy or strategic intent.

Additionally, while studies of political trolls frequently distinguish between domestic and foreign actors, particularly in the context of coordinated influence campaigns \cite{blinder_ware_2024_sweden_troll_farm}, we do not classify commenters by origin. Our analysis centers on behavioral manifestations of incivility toward partisan out-groups, rather than assumptions about identity, motivation, or geographic provenance.

Crossing into hostile territory is rare in selective-exposure environments, yet those who do so often embody a confrontational form of partisan expression. Similar patterns are noted in prior work on cross-partisan engagement on YouTube, where conservatives were more likely to comment on liberal channels \cite{wu_cross-partisan_2021}. Our contribution is to formalize this behavior with the concept of partisan warriors and measure its prevalence across different types of commenters, moving from isolated cases to a systematic framework for identifying targeted incivility.

\subsection{Video Content, Toxicity, and Commenters}
Channels are not neutral backdrops as the videos they host shape the discursive climate and the commenters they attract. Prior research has demonstrated that content frames shape political attitudes and conversational tone. Studies of YouTube in particular find that algorithmic curation and radicalizing content can steer users toward more extreme forms of engagement \cite{ribeiro_auditing_2020}. Toxic or inflammatory videos create an environment that may embolden hostile participation while deterring deliberative voices.

\citet{jiang_bias_2019} demonstrate that partisanship and misinformation on YouTube interact with commenting behaviors, reinforcing divides. Nevertheless, there is a dearth of work systematically connecting video toxicity to the composition of commenters. Understanding this link is essential for explaining why some channels attract partisan warriors more than others. We extend this line of research by linking video-level toxicity to the presence of partisan warriors in comment sections, leveraging LLMs to scale and nuance toxicity detection at the content level.

\section{Data and Methods}
\subsection{Data Collection}
\subsubsection{Channel Selection and Classification}
Our study leveraged the US Partisan Media Dataset developed by \citet{wu_cross-partisan_2021}, which provides a comprehensive classification of YouTube channels based on their political orientation. From this dataset, we selected channels that had been previously classified by Media Bias/Fact Check (MBFC)\footnote{\url{https://mediabiasfactcheck.com/}} as either left-leaning or right-leaning sources. The final sample comprised 551 channels, with 293 classified as left-leaning and 258 as right-leaning.

\subsubsection{Video Data Collection}
Video data were collected using the YouTube Data API V3 during a strategically selected timeframe surrounding the second U.S. presidential debate of 2024 (September 10, 2024). We focused on the week immediately following the debate because such high-profile political events act as a common external shock, triggering heightened engagement and rapid shifts in discourse. This window provides a unique opportunity to examine how commenters' behaviors--such as patterns of toxicity and attack targets--evolve in response to the same political stimulus. To ensure content relevance, videos were included only if they contained ``Trump'' or ``Harris'' in their titles or descriptions, directly linking the content to the two main candidates. This approach yielded a total of 2,920 videos across 160 channels (86 left-leaning and 74 right-leaning channels), with left-leaning channels contributing 1,710 videos and right-leaning channels contributing 1,210 videos.

\begin{table}[ht]
\centering
\begin{tabular}{lrrr}
\toprule
 & \textbf{Total} & \makecell{\textbf{On liberal}\\\textbf{channels}} & \makecell{\textbf{On conservative}\\\textbf{channels}} \\
\midrule
Channels & 160      & 86       & 74 \\
Videos   & 2,920    & 1,710    & 1,210 \\
Comments & 1,854,320 & 1,294,519 & 559,801 \\
\bottomrule
\end{tabular}
\caption{Dataset overview by channel leaning.}
\label{tab:dataset}
\end{table}

\subsubsection{Comment Data Collection}
We collected comment data using the YouTube API during a strategically defined temporal window to capture immediate reactions and discussions surrounding the presidential debate. The collection period spanned two weeks (September 10-24, 2024), beginning with the debate itself and extending through the subsequent week when related videos were published and comments actively accumulated. This timeframe ensured comprehensive coverage: we captured both debate-related videos uploaded in the week following the event and allowed an additional week for viewer engagement. This approach is particularly important as early commenters often play a disproportionate role in shaping video discourse and represent the most politically engaged participants \cite{xiao_changing_2019, tan_winning_2016, miyazaki_impact_2024}. By restricting our window to this two-week period, we focused on the most salient and immediate forms of electoral engagement.

We conducted data retrieval retrospectively between October 26 and October 28, 2024, approximately one month after the defined analysis window. Prior research on moderation bias suggests that although moderation decisions are generally unbiased when controlling for hate speech and extremeness, certain political spaces exhibit a higher prevalence of toxic attributes, which can lead to disproportionate removal rates for conservative content \cite{jiang_reasoning_2020}. Although the one-month gap between the analysis window and data retrieval could have been influenced by platform moderation and channel-level curation, our dataset reflects the ``surviving'' discourse--namely, comments that remained visible at the time of collection.

For each video, we collected only top-level (root) comments that are not replies to other comments; these reflect viewers' initial reactions to the video content and serve as the starting points of discussion threads. The final dataset comprised 1,854,320 comments from 677,238 unique commenters. As shown in Table \ref{tab:dataset}, 1,294,519 comments were written on videos from liberal channels, while 559,801 comments were written on videos from conservative channels. While this disparity partly reflects differences in the number of videos per channel type, it is also influenced by the differential moderation dynamics described above. Consequently, our analysis likely provides a conservative, lower-bound estimate of the true extent of toxicity and partisan hostility, as the most extreme instances were presumably removed prior to our data collection.

\subsection{Cross-Cutting Commenter Identification}

\begin{table}[ht]
\centering
\begin{tabular}{lrr}
\toprule
\textbf{Commenter Type} & \textbf{Commenters} & \textbf{Comments} \\
\midrule
Only-Liberal       & 411,893 (60.8\%) & 919,376 (49.6\%) \\
Only-Conservative  & 187,365 (27.7\%) & 323,212 (17.4\%) \\
Cross-Cutting      &  77,980 (11.5\%) & 611,732 (33.0\%) \\
\midrule
\textbf{Total}      & 677,238  & 1,854,320 \\
\bottomrule
\end{tabular}
\caption{Distribution of commenters and comments by commenter type.}
\label{tab:distribution_comment}
\end{table}

To categorize commenters, we followed prior work that classifies users based on their activity across ideologically distinct channels \cite{han_news_2023}. We distinguished between liberal-only commenters (who posted comments exclusively on liberal-leaning channels), conservative-only commenters (who posted comments exclusively on conservative-leaning channels), and cross-cutting commenters (who posted comments on both channels). Table~\ref{tab:distribution_comment} shows the distribution of commenters and comments by type. Notably, although cross-cutting commenters represented just 11.5\% of all individuals, they produced fully one-third of all comments, underscoring their outsized role in shaping discourse across partisan boundaries. To further assess the robustness of this categorical classification, we also examined a continuous measure of cross-ideological engagement based on the fraction of comments posted on liberal-leaning channels; the results, reported in Appendix, support the validity of our grouping strategy.

\subsection{Detecting Attack Targets in Comments using LLMs}

\subsubsection{Data Filtering Pipeline}
To focus our analysis on persistent patterns of political incivility, we applied a two-step filtering pipeline to the raw dataset. 

First, we isolated comments most likely to contain hostile or derogatory language by measuring toxicity using Perspective API’s toxicity model \cite{10.1145/3534678.3539147}. Consistent with prior work on political incivility \cite{gervais_incivility_2025}, we applied a 90th percentile threshold ($\geq$ 0.578) to retain only the most uncivil subset of discourse.

Second, to ensure our analysis captured behavioral patterns rather than isolated incidents, we restricted the sample to \textit{active users}, defined as those who posted three or more comments that met the toxicity threshold. This step is crucial for establishing a reliable user-level stance in subsequent sections; a minimum of three data points prevents the classification of users based on a single, potentially anomalous outlier \cite{wu_cross-partisan_2021}. 

The intersection of these two filters--high toxicity and repeated engagement--yielded a final dataset of 62,598 comments for attack target analysis. While this excludes non-toxic comments and one-off posters, this focused sampling strategy is necessary to reliably model the specific phenomenon of sustained partisan hostility.

\begin{table*}[h]
\centering
\begin{tabular}{p{0.25\textwidth} p{0.12\textwidth} p{0.10\textwidth} p{0.25\textwidth} p{0.16\textwidth}}
\toprule
\textbf{Comment} & \textbf{Channel Name} & \textbf{Channel Leaning} & \textbf{Video Title} & \textbf{Predicted Attack Target(s)} \\
\midrule
``Hey Trump, don’t you dare grab that hag’s hand''
& Fox News 
& Conservative 
& Doocy: Kamala Harris has some explaining to do 
& Kamala Harris \\
\midrule
``Notice they always used `apparent assassination attempt' because he was a radical left Biden/Harris supporter brainwashed by the radical far left media garbage propaganda''
& MSNBC 
& Liberal 
& Journalist who interviewed suspect of apparent Trump assassination attempt last year speaks out
& Supporters of Kamala Harris, Mainstream media \\
\midrule
``Keep your mouth shut, puppet!''
& The Washington Examiner 
& Conservative
& ``I'm Talking Now'': Trump Quotes Harris After She Interrupts \#shorts
& Unclear \\
\midrule
``Claiming Haitians eat people’s pets is basically calling them less than human. DonOld is utterly deranged. Maybe if he stopped binging right-wing trash TV, he wouldn’t come off sounding like such an idiot.''
& CNN 
& Liberal
& ``Knock It Off'': Ohio AG Reacts to Uptick in Threats After JD Vance's False Claims About Immigrants
& Donald Trump, Mainstream Media \\
\bottomrule
\end{tabular}
\caption{Examples of attack target classification of comment.}
\label{tab:example}
\end{table*}

\subsubsection{Model Selection}
We employed GPT-4.1, a state-of-the-art large language model, to identify attack targets in each comment. LLMs offer advantages for this task given their ability to process contextual nuance at scale, particularly important for analyzing contemporary political discourse \cite{heseltine_large_2024}.

\subsubsection{Taxonomy Development Prompt Design}
To capture the nuances of political incivility during the election cycle, we developed an 18-category taxonomy that extends \citet{rossini_more_2021}'s framework. The taxonomy was designed to reflect the personality-driven nature of contemporary U.S. politics and to distinguish between attacks on political leaders, their supporters, and broader groups. Categories include individual political actors (e.g., Donald Trump, Kamala Harris, Joe Biden), political groups (e.g., Republicans, Democrats, conservatives, liberals), protected or marginalized groups (e.g., immigrants, racial or gender identity groups, religious communities), institutional actors (e.g., debate moderators, mainstream media, government bodies), and an \textit{Unclear/Other} category for residual cases.  To implement this taxonomy, we prompted GPT-4.1 to classify only the \textit{Attack target(s)}--entities explicitly or implicitly addressed with uncivil language--while excluding civil criticism or neutral mentions. Also, we adopted a conservative contextual resolution strategy for vague references (e.g., ``he,'' ``they'') \cite{gan-etal-2024-assessing}. The LLM was permitted to consult video metadata only when strong evidence supported a clear resolution; otherwise, the reference was coded as \textit{Unclear}. This procedure ensured consistent operationalization of the taxonomy while avoiding over-interpretation. Examples of comment attack target classifications can be found in Table \ref{tab:example}, and the GPT-4.1 prompt example used for attack target detection is provided in the Appendix.

\subsubsection{Validation}
To establish reliability, two authors trained in political communication independently coded 100 comments, yielding a Jaccard similarity of 0.865, indicating strong human agreement. The Jaccard Index was chosen because it is well-suited for multi-label classification tasks, where comments may include multiple targets \cite{pmlr-v63-Gouk8, levandowsky_distance_1971}. We then compared human and LLM outputs on a stratified sample of 626 comments (1\% of the dataset), testing temperature values from 0.1 to 0.5 with five iterations each. Agreement peaked at temperature 0.3 (\textit{Jaccard} = 0.724), demonstrating substantial human–LLM alignment.

\paragraph{Final Setup.}
Based on these results, we fixed the decoding temperature at 0.3 for all subsequent classifications, ensuring an optimal trade-off between consistency and contextual flexibility. This validation confirms that our LLM-based approach provides a scalable and reliable method for mapping incivility targets in large-scale political discussions. As shown in Appendix Figure~\ref{fig:toxicity_all}, the distribution of attack targets also varies by commenter type depending on the channel in which they posted.

\subsection{Operationalizing Partisan Warriors}

\subsubsection{Stance Classification}
\begin{table}[h]
\centering
\begin{tabular}{lccc}
\toprule
\multirow{2}{*}{\textbf{Commenter Type}} & \multicolumn{3}{c}{\textbf{Stance Label(\%)}} \\
\cmidrule(l){2-4}
 & Anti-Lib. & Anti-Con. & Unclear \\
\midrule
Only-Liberal & 6.9 & 92.5 & 0.6 \\
Only-Conservative & 69.2 & 30.0 & 0.8 \\
Cross-cutting & 29.3 & 69.6 & 1.1 \\
\bottomrule
\end{tabular}
\begin{flushleft}
\end{flushleft}
\caption{Cross-tabulation of commenter type and inferred stance.}
\label{tab:stance_validity}
\end{table}

We first classified commenters according to the partisan alignment of their uncivil attacks. Each attack target identified by the LLM was mapped into an ideological polarity. Categories referencing Supporters of Donald Trump, Donald Trump, Republicans, or conservatives in general (IDs 1, 4, 5, and 8) were coded as \textbf{anti-conservative attacks}, while those referencing Supporters of Kamala Harris, Kamala Harris, Democrats, or liberals in general (IDs 2, 3, 6, and 9) were coded as \textbf{anti-liberal attacks}. Each comment could increment both dimensions if multiple categories were targeted simultaneously. These values were then aggregated at the commenter level, producing for each user the total count of anti-liberal ($A^{(l)}_u$) and anti-conservative ($A^{(c)}_u$) attacks.

To quantify the partisan orientation of a user's hostility, we calculated the \textbf{Stance Score}. This metric represents the normalized difference between a user's attacks on liberals versus conservatives:
\[
\text{stance\_score}_u = \frac{A^{(l)}_u - A^{(c)}_u}{A^{(l)}_u + A^{(c)}_u}
\]
where $A^{(l)}_u$ and $A^{(c)}_u$ represent the total number of anti-liberal and anti-conservative attacks by user $u$, respectively. The score ranges from $-1$ (exclusively attacking conservatives) to $+1$ (exclusively attacking liberals), with values near zero reflecting mixed or balanced attack behavior. Commenters who produced no attacks were assigned missing values.

For categorical analysis, we discretized the \text{stance\_score} into three labels: \textit{Anti-Liberal} (\text{stance\_score} $\geq 0.2$), \textit{Anti-Conservative} (\text{stance\_score} $\leq -0.2$), and \textit{Unclear} ($-0.2 < \text{stance\_score} < 0.2$). The $\pm 0.2$ threshold was selected based on sensitivity analysis spanning 0.2 to 0.8 (see Appendix Table \ref{tab:threshold}), ensuring that labeled users exhibited a distinct direction of hostility rather than a balanced mix of attacks.

\subsubsection{Partisan Warrior Classification}
We operationalized \textbf{Partisan Warriors (PWs)} as users who enter ideologically opposing spaces to attack that group (e.g., an anti-liberal user attacking liberals within a liberal channel).

Unlike general partisan commenters who may vent frustration within their own echo chambers, PWs are defined by the intersection of their \textbf{Inferred Stance} and their \textbf{Commenting Arena}. A user was classified as a PW if they met the following criteria:

\begin{itemize} 
    \item \textbf{Directional Hostility:} The user must have a clear stance (either Anti-Liberal or Anti-Conservative).
    \item \textbf{Oppositional Context:} The user must deploy these attacks in an environment ideologically aligned with their target.
\end{itemize}

Based on their channel activity, users were classified as follows:

\begin{itemize} 
    \item \textbf{Single-Side Commenters:} Users active on only one side of the media ecosystem were labeled PWs only if they engaged in counter-attitudinal attacks (i.e., invading the opposing camp).
    \begin{itemize} 
        \item Users on \textit{liberal channels} were labeled PWs if they had an \textbf{Anti-Liberal} stance. 
        \item Users on \textit{conservative channels} were labeled PWs if they had an \textbf{Anti-Conservative} stance. 
    \end{itemize} 
    \item \textbf{Cross-Cutting Commenters:} Users active on both liberal and conservative channels were labeled PWs if they possessed \textit{any} distinct stance.
\end{itemize}

\subsection{Analytical Overview by Research Question}
Because each research question targets a distinct analytical goal, the unit of analysis and the subset of data used vary across RQ1--RQ4. We provide a visual overview of this methodological pipeline in Appendix Figure \ref{fig:pipeline}.

RQ1 and RQ2 investigate broad patterns of incivility and audience response without relying on the specific Partisan Warrior (PW) classification. RQ1 provides a baseline comparison of incivility across commenter types using toxicity scores for \emph{all collected comments}. RQ2 examines audience endorsement mechanisms, maintaining the \emph{individual comment} as the unit of analysis. Here, we model the number of likes received by targeted toxic comments as a function of their attack targets, retaining comment-level granularity to capture variation in audience rewards.

In contrast, RQ3 and RQ4 specifically analyze the PW phenotype, requiring aggregated user-level or interaction-level data. RQ3 focuses on the prevalence of PWs, necessitating a \emph{commenter-level analysis} ($N = 12{,}884$ unique commenters). For this inquiry, we aggregate comments posted by active commenters to classify each unique commenter based on their overall cross-posting behavior and stance, comparing the proportion of identified PWs across different commenter groups.

Finally, RQ4 investigates environmental triggers for PW participation using a dataset of \emph{unique commenter-within-video} observations ($N = 57{,}644$). To ensure statistical independence and avoid pseudoreplication from hyper-active users, if a user posted multiple comments on the same video, their activity was treated as a single participation event. Using this deduplicated dataset, we employ multilevel modeling to estimate the likelihood that a participating commenter was classified as a PW, with video-level properties (e.g., title toxicity) as predictors and random intercepts to account for clustering within videos and channels.

\section{Results}
\subsection{RQ1: Incivility of Cross-cutting vs. Single-side Commenters}
\begin{figure}[h]
\centering
\includegraphics[width=0.95\columnwidth]{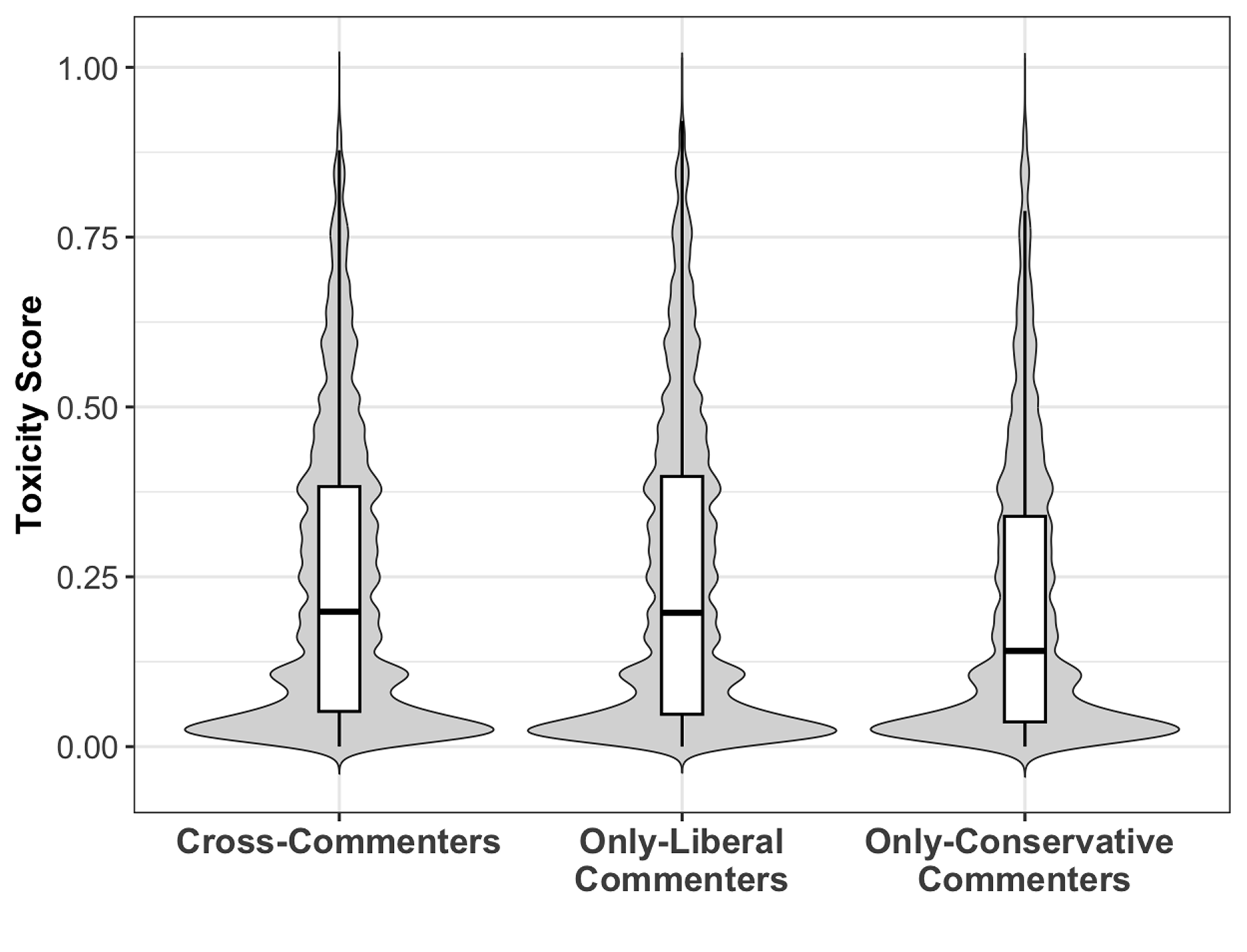} 
\caption{Distribution of toxicity scores by commenter group. Violin plots show the full distribution, with boxplots overlaid for medians and interquartile ranges. Cross-cutting commenters are active on both liberal- and conservative-leaning channels, while only-liberal and only-conservative commenters denote activity restricted to liberal- or conservative-leaning channels. While the overall test indicates significant differences across groups, post-hoc comparisons show no significant difference in toxicity between cross-cutting and only-liberal commenters.}

\label{fig:toxicity_all}
\end{figure}

To address RQ1, we examined whether cross-cutting commenters differ in toxicity from single-side commenters who comment only on liberal-leaning or only on conservative-leaning channels. The average toxicity scores were similar for cross-cutting ($M = 0.247$, $SD = 0.215$) and liberal-only commenters ($M = 0.252$, $SD = 0.225$), whereas conservative-only commenters showed a lower mean toxicity ($M = 0.215$, $SD = 0.210$).

Given the skewed distribution of toxicity, we employed a Kruskal–Wallis test, which indicated significant differences among the three groups, $\chi^{2}(2) = 7{,}203.4, p < .001$. Post-hoc Dunn’s tests with Bonferroni correction showed no significant difference between cross-cutting and liberal-only commenters ($p = .32$), but both groups were significantly more toxic than conservative-only commenters ($p < .001$ for both comparisons). Taken together, these results indicate that cross-cutting commenters are neither more civil nor more uncivil than liberal-only commenters, while both cross-cutting and liberal-only commenters exhibit higher toxicity than conservative-only commenters.

To interpret these scores, it is helpful to examine representative examples. Comments near the conservative-only mean ($0.21$) often featured slogan-based dismissals or rhetorical sarcasm, such as \textit{``What spice goes best with liberal tears?''} or \textit{``Klamedia obviously doesn't care! Never vote Democrat!''} Similarly, comments near the cross-cutting mean ($0.25$) frequently employed descriptive criticism or mockery, such as \textit{``Trump is a fugitive from the law''} or \textit{``Send him to Mars with Elon Musk ... They can keep themselves busy stroking their egos.''} 

These examples suggest that, despite statistically significant differences in average toxicity, the substantive differences in linguistic intensity across groups are relatively modest. Rather than reflecting sharp contrasts in severity, toxicity scores in this range appear to capture broadly similar forms of partisan hostility expressed through different rhetorical styles. This finding highlights a key limitation of relying solely on aggregate toxicity scores: such measures do not distinguish between different behavioral orientations underlying hostile speech. In other words, it is not necessarily the \textit{level} of toxicity that differentiates engagement patterns, but potentially \textit{who} is being targeted. 

\subsection{RQ2: Asymmetric Audience Rewards for Incivility}

\begin{figure}[t]
    \centering
    \includegraphics[width=\linewidth]{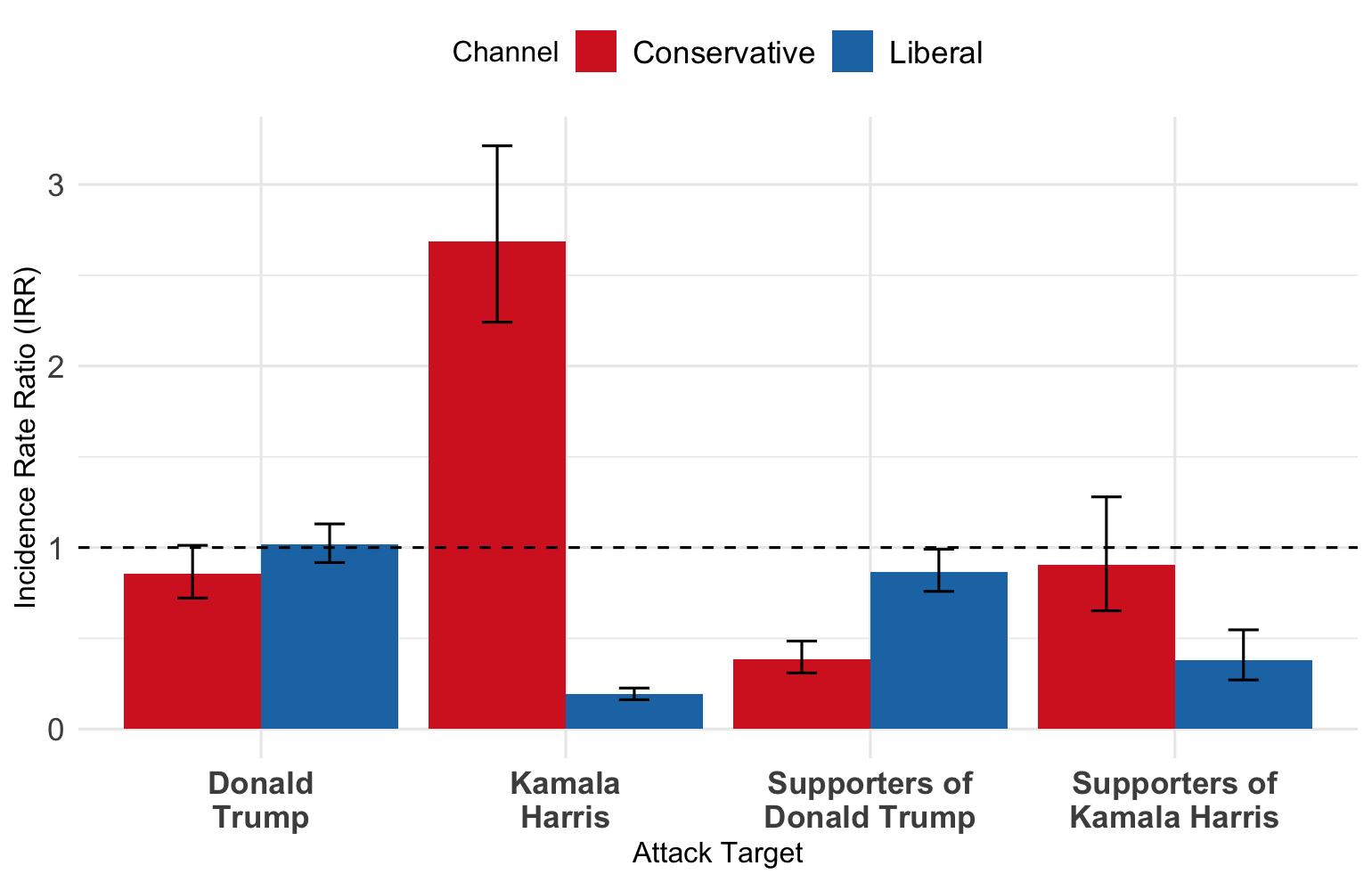}
    \caption{Incidence Rate Ratios of audience likes for uncivil comments across targets and channel leanings, estimated via negative binomial regression. Error bars indicate 95\% confidence intervals.}
    \label{fig:irr}
\end{figure}

To examine whether the targets of uncivil comments influenced audience endorsement, we modeled the number of likes received by each uncivil comment using negative binomial regression, which is appropriate for over-dispersed count outcomes. Separate models were fit for liberal-leaning and conservative-leaning channels. In each model, the dependent variable was the count of likes, and the key predictor was the attack target category. The reference category was set to \textit{Unclear} targets. We report incidence rate ratios (IRRs), obtained by exponentiating the model coefficients, with 95\% confidence intervals (Figure \ref{fig:irr}).

On liberal channels, results indicate that uncivil comments directed at \textit{Supporters of Donald Trump} received significantly fewer likes than unclear attacks (IRR = 0.87, $p = .036$). Comments targeting \textit{Supporters of Kamala Harris} were even less endorsed (IRR = 0.38, $p < .001$), and those directly attacking \textit{Kamala Harris} received the least support (IRR = 0.19, $p < .001$), amounting to an 81\% reduction relative to the unclear baseline. In contrast, uncivil comments targeting \textit{Donald Trump} obtained a similar level of likes as unclear attacks (IRR = 1.02, $p = .72$). These results suggest that left-leaning audiences systematically withdraw support from uncivil comments aimed at ingroup figures, while showing no significant preference between attacking Trump versus issuing vague, non-specific attacks.

A distinctively different pattern emerged on conservative channels. Uncivil comments targeting \textit{Supporters of Donald Trump} were strongly disfavored (IRR = 0.39, $p < .001$), attracting only about 39\% of the likes received by unclear attacks. By contrast, uncivil comments directed at \textit{Kamala Harris} were highly rewarded, receiving nearly 2.7 times as many likes as the unclear baseline (IRR = 2.69, $p < .001$). Attacks on \textit{Supporters of Kamala Harris} (IRR = 0.90, $p = .55$) and \textit{Donald Trump} himself (IRR = 0.86, $p = .072$) did not significantly differ from unclear attacks.

Thus, right-leaning audiences strongly endorse uncivil comments against Harris while penalizing attacks on their own ingroup supporters. Together, these findings reveal that the relationship between uncivil discourse and audience endorsement is contingent on both the ideological leaning of the channel and the specific attack target. This asymmetric incentive structure--where out-group attacks are rewarded in some spaces but not others--provides the environmental context for analyzing the prevalence of \textit{Partisan Warriors} in the subsequent section (RQ3).

\subsection{RQ3: Partisan Warrior Prevalence by Commenter Type}

To examine whether PW tendencies vary across commenter types, we conducted a chi-squared test of independence between commenter type (Cross-cutting Commenters (Cross-Commenters), Only-Liberal Commenters, Only-Conservative Commenters) and PW status at the unique commenter level ($N = 12,884$). Cross-Commenters are active on both liberal- and conservative-leaning channels, while Only-Liberal Commenters and Only-Conservative Commenters denote activity restricted to liberal- or conservative-leaning channels. 

The distribution differed substantially across groups, $\chi^2(2) = 9{,}222.1, p < .001$, with a large effect size (Cramér’s $V = .85$). As expected given our operational definition, nearly all Cross-Commenters were classified as PWs (93.6\%). In contrast, PW proportions diverged sharply between single-leaning commenters: only 6.6\% of Only-Liberal Commenters were PWs, whereas 27.3\% of Only-Conservative Commenters fell into this category.

To further quantify these differences, we ran a logistic regression predicting PW status from commenter type (reference: Only-Conservative Commenters). Results showed that Cross-Commenters had markedly higher odds of being PWs (OR = 38.80, 95\% CI [32.6, 46.4], $p < .001$), while Only-Liberal Commenters had dramatically lower odds (OR = 0.19, 95\% CI [0.16, 0.22], $p < .001$). Model performance was strong ($R^2_{\text{Tjur}} = .72$), indicating that commenter type alone accounted for a large portion of variance in PW behavior.

Taken together, these results confirm that Cross-Commenters overwhelmingly meet the partisan warrior definition, which is consistent with the operationalization itself. More importantly, we observe a striking asymmetry between the two single-channel groups. While the majority of Only-Conservative Commenters were non-warriors, they were more than four times as likely to engage in partisan warrior behavior compared to their Only-Liberal counterparts (27.3\% vs. 6.6\%). This imbalance highlights that PW activity is not evenly distributed across ideological camps but is disproportionately concentrated among commenters who primarily engage on conservative channels.

\begin{table}[h]
\centering
\label{tab:rq3_dist}
\begin{tabular}{lrr}
\toprule
\textbf{Commenter Type} & \textbf{Not PW} & \textbf{PW} \\
\midrule
Cross-cutting & 380 (6.4\%) & 5,534 (93.6\%) \\
Only-Liberal   & 5,627 (93.4\%) & 397 (6.6\%) \\
Only-Conservative   & 688 (72.7\%) & 258 (27.3\%) \\
\midrule
\textbf{Total}   & 6,695 & 6,189 \\
\bottomrule
\end{tabular}
\caption{Distribution of PW classification by commenter type (Unique commenter-level analysis).}
\end{table}

\subsection{RQ4: Effects of Video Title Toxicity on Partisan Warrior Participation}

\begin{figure}[h]
\centering
\includegraphics[width=0.9\columnwidth]{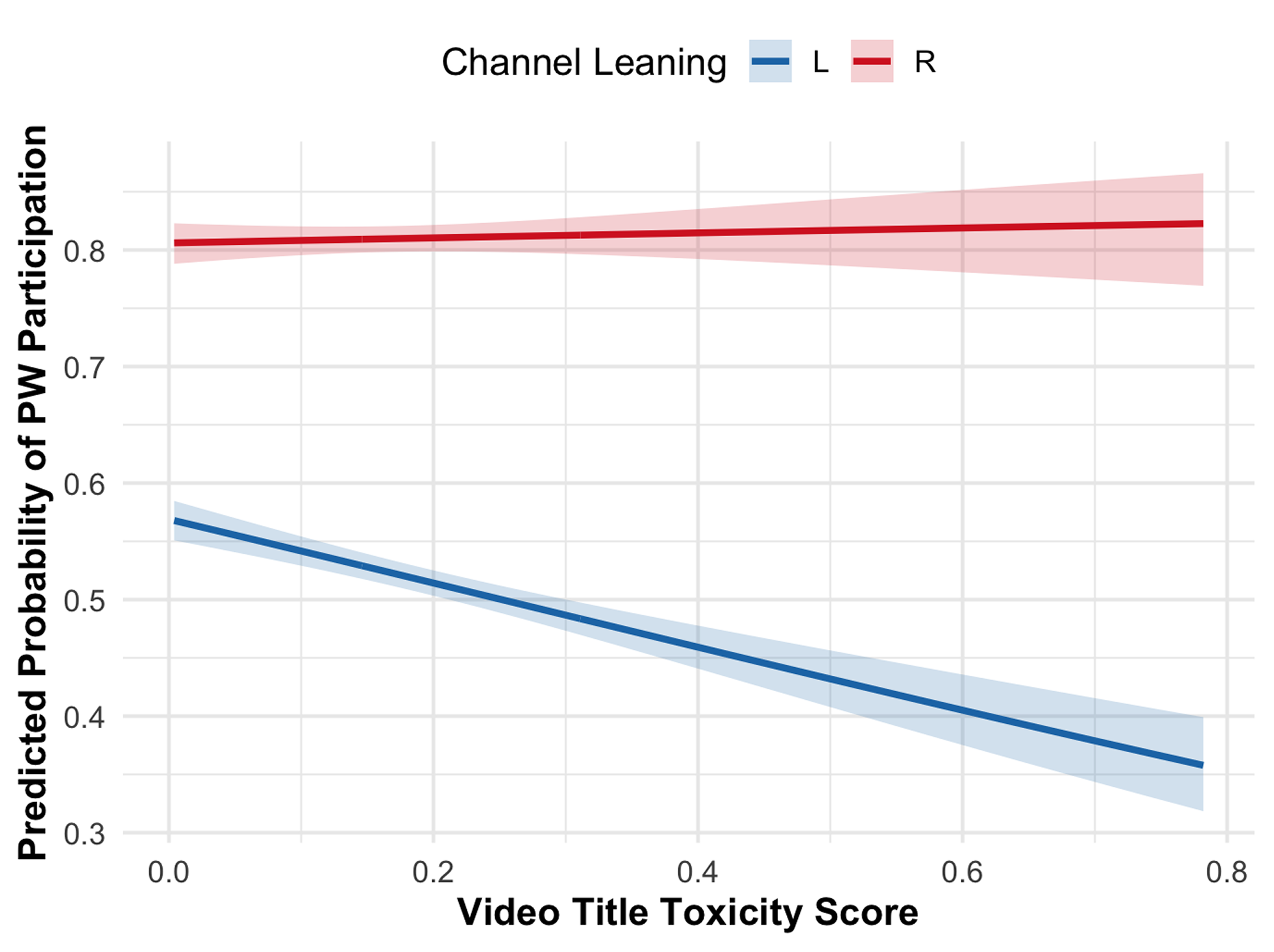} 
    \caption{Predicted probabilities of PW participation by video title toxicity and channel leaning, based on the Baseline Model. Shaded regions represent 95\% confidence intervals. The plot illustrates the apparent ideological asymmetry: toxic titles appear to deter PWs on liberal channels (blue) but have a negligible or slightly mobilizing effect on conservative channels (red).}
    \label{fig:rq4_interaction}
\end{figure}

Given the disproportionate concentration of Partisan Warriors in conservative spaces identified in RQ3, we next investigate the environmental triggers that might drive this engagement. Specifically, to examine whether toxic video titles are associated with the likelihood of PW participation, we fitted a series of generalized linear mixed models (GLMMs) with a binomial link function using the dataset of commenter-video pairs (N = 57,644). Descriptive statistics within this dataset highlight a baseline asymmetry: while 79.3\% of participation events on conservative channels involved PWs, this rate was markedly lower on liberal channels (45.1\%). In contrast, the average toxicity of video titles was nearly identical across channel ideologies (liberal: $M = 0.164$, $SD = 0.143$; conservative: $M = 0.167$, $SD = 0.141$), suggesting that aggregate title framing does not differ systematically by political leaning.

\subsubsection{Baseline Association}
We first estimated a baseline model including a random intercept for each video to account for unobserved video-level heterogeneity ($AIC = 72{,}531$).  In this model, video title toxicity was negatively associated with PW participation on liberal-leaning channels (\textit{Estimate} = $-1.10$, $SE = 0.15$, $p < .001$), suggesting that toxic titles deterred PW participation. Conversely, the interaction term between title toxicity and channel leaning was positive and significant (\textit{Estimate} = $1.24$, $SE = 0.30$, $p < .001$). This indicated a strong ideological asymmetry: while toxic titles appeared to suppress PW activity on liberal channels, this effect was effectively offset on conservative channels, suggesting a potential mobilization effect in the baseline specification.

\subsubsection{Robustness Check: Channel Identity}
However, considering that videos are embedded within channels that possess distinct community norms and audience bases, we conducted a robustness check to determine whether these effects were driven by immediate video cues or stable channel characteristics. We extended the model by adding a random intercept for \textit{channel identity} to control for channel-level heterogeneity.

\begin{figure}[h]
    \centering
    \includegraphics[width=1.0\columnwidth]{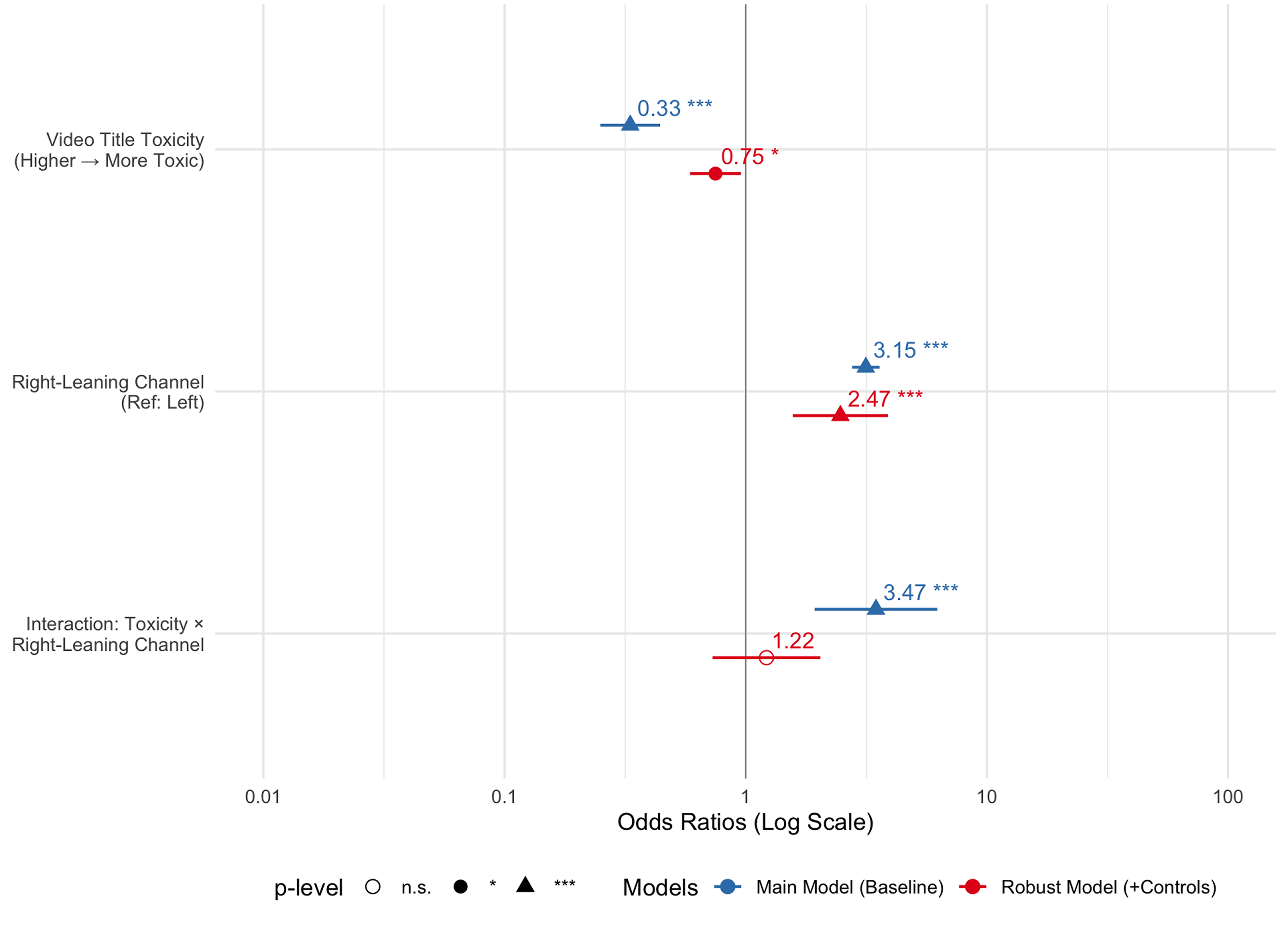} 
    \caption{Comparison of fixed effects between the Main Model (Baseline, Blue) and the Robust Model (Red). Error bars represent 95\% confidence intervals. Notably, the interaction term is significant in the Baseline Model but becomes non-significant in the Robust Model (indicated by the hollow circle), confirming that the apparent asymmetry is driven by channel-level heterogeneity.}
    \label{fig:rq4_coef_comparison}
\end{figure}

The extended model demonstrated a significantly better fit to the data ($\Delta AIC = 547$, $\chi^2(1) = 548.5$, $p < .001$), indicating that channel-level variation is a critical predictor of PW behavior.

Crucially, after controlling for channel identity, the ideological asymmetry observed in the baseline model vanished. The interaction between title toxicity and channel leaning was no longer statistically significant (\textit{Estimate} = $0.20$, $SE = 0.26$, $p = .450$). This implies that the differential response to toxicity observed in the baseline model was an artifact of channel composition rather than a true interaction with video framing.

Furthermore, the main effect of title toxicity remained negative and statistically significant, though substantially weaker in magnitude (\textit{Estimate} = $-0.29$, $SE = 0.12$, $p = .020$). Within channels, higher video title toxicity is associated with a modest decrease in PW participation, indicating that toxic framing does not mobilize PWs once stable channel norms are taken into account.

\subsubsection{Summary}
These findings suggest that PW participation is primarily an ecological phenomenon structured at the channel level. The apparent association between toxic titles and PW activity on conservative channels is explained by the fact that PWs tend to inhabit specific channels that have higher baseline levels of engagement. Once the stable channel context is accounted for, the specific framing of an individual video title has minimal impact on mobilizing Partisan Warriors. Thus, PW behavior appears to be governed more by established community norms and channel ecosystems than by transient content triggers.

\section{Conclusion and Discussion}

This study, grounded in the immediate aftermath of the second 2024 U.S. presidential debate, offers a granular examination of the conditions under which partisan hostility unfolds on YouTube. By shifting the analytical lens from aggregate toxicity to the directionality and context of attacks, we challenge existing assumptions about cross-cutting interaction and reveal the structural dynamics that sustain ``Partisan Warrior'' behavior.

First, our investigation into cross-cutting commenters challenges the idealized view that traversing ideological divides fosters deliberation (RQ1). We found that these users were not distinctively civil; rather, their toxicity levels were indistinguishable from liberal-only commenters and significantly higher than conservative-only commenters. This pattern should be interpreted with caution, as the data reflect post-moderation content collected approximately one month after the analysis window, and prior work suggests that moderation may differentially affect the visibility of toxic comments across political spaces \cite{jiang_reasoning_2020}. 

At the same time, the qualitative similarity in rhetorical intensity across groups underscores the critical limitations of relying solely on aggregate toxicity scores. In a polarized landscape where ideological mobility often signals hostility rather than dialogue, generic metrics fail to distinguish between internal policing and out-group aggression. This motivated our shift toward operationalizing the Partisan Warrior construct, which defines users not merely by how toxic they are, but by whom their attacks target.

Examining the surrounding environment helps explain why such behavior persists (RQ2). Our analysis of audience endorsement revealed an asymmetry in incentive structures. While audiences in both liberal and conservative spaces penalized in-group attacks--reflecting a shared preference for internal cohesion--the rewards for out-group hostility diverged sharply. On conservative channels, uncivil attacks targeting Kamala Harris received nearly three times as many likes as baseline uncivil comments, forming a potent engagement loop that validates hostile behavior. No comparable reward structure emerged on liberal channels, where attacks on Donald Trump received no significant engagement boost. This pattern suggests that the online ``marketplace of ideas'' is not neutral: hostility toward the out-group is socially reinforced in conservative spaces, whereas comparable behavior in liberal spaces lacks similar validation.

Taken together, these incentive asymmetries align with the structural differences in Partisan Warrior prevalence across ideological spaces (RQ3). Conservative channels appear to function as contested battlegrounds, characterized by frequent hostile incursions and strong social rewards for retaliatory behavior. In these environments, resident users are disproportionately likely to adopt Partisan Warrior tactics; indeed, only-conservative users exhibited a Partisan Warrior rate (27.3\%) more than four times higher than their only-liberal counterparts (6.6\%). We interpret this pattern as consistent with an ecological dynamic in which environments marked by higher perceived antagonism and asymmetric rewards are more likely to sustain resident hostility. By contrast, liberal spaces more closely resemble homogenous safe havens, where lower levels of perceived threat coincide with weaker incentives for out-group aggression and a minimal presence of resident warriors.

Finally, our analysis of environmental triggers clarifies the locus of this hostility (RQ4). Although baseline models suggested a link between video title toxicity and PW participation, robustness checks revealed that these patterns were largely explained by channel-level heterogeneity. This indicates that Partisan Warrior behavior is not a transient reaction to immediate framing cues (e.g., a toxic video title) but an ecological phenomenon driven by channel-specific norms. This interpretation aligns with prior work showing that ideological concentration on social media is driven more by user self-selection and network homophily than by algorithmic ranking or content-level triggers \cite{bakshy_exposure_2015}. Moreover, research on online moderation demonstrates that toxic behavior is not solely an individual attribute but is strongly conditioned by the norms of the communities in which users participate \cite{chandrasekharan_you_2017}. Given evidence that users entering political communities rapidly adapt their behavior to local norms \cite{rajadesingan_quick_2020}, the observed asymmetry in PW participation is best understood as emerging from enduring, channel-specific ecosystems--shaped by the incentive structures identified in RQ2--rather than from the framing of individual videos.

These findings extend the understanding of online political discourse by demonstrating that partisan hostility is not a uniform feature of the internet but a phenomenon structured by specific environmental incentives. For platform governance, this implies that interventions must move beyond content-level policing to address ecosystem-level dynamics.

First, on the algorithmic side, ranking systems that may inadvertently gamify hostility--by rewarding or amplifying comments that attack out-groups--require recalibration. Downranking uncivil comments that disproportionately target partisan opponents could dismantle the reward loops that fuel partisan warrior dynamics. Second, given that hostility is rooted in channel ecosystems (RQ3, RQ4), moderation policies should focus on identifying and prioritizing channels that foster hostile norms for ecosystem-level interventions, rather than merely flagging individual assets. Ultimately, mitigating partisan warfare requires reshaping the structural ``battlegrounds'' and incentive systems that allow such behavior to flourish.

\subsection{Ethical Statements}
This research adheres to established ethical guidelines for social media research. We analyzed only publicly available comments collected through YouTube’s official API, in full compliance with its terms of service. To protect user privacy, all identifiers were anonymized, and the analysis focused on aggregate patterns rather than individuals. Following the Association of Internet Researchers (AoIR) guidelines, all example quotes in Table \ref{tab:example} were paraphrased to preserve meaning while ensuring anonymity and preventing traceability.

We recognize that, as with any study of partisan dynamics, findings could be misrepresented or selectively cited to intensify political conflict. For example, the asymmetry in partisan warrior behavior might be oversimplified to stigmatize one side, or insights about environmental triggers misused to design manipulative content. However, the intention of this work is explicitly diagnostic and constructive. Our analysis aims to illuminate mechanisms of online hostility so that platforms, policymakers, and scholars can better understand and mitigate polarization. By foregrounding methodological transparency and careful interpretation, we seek to minimize misuse while maximizing the potential to inform healthier and more inclusive online political dialogue.

\subsection{Limitations and Future Work}
While this study provides insights into the asymmetric nature of online partisan hostility, several limitations should be noted, opening avenues for future research. First, our analysis is restricted to YouTube and a curated set of liberal and conservative channels in the U.S., focusing on comments after the second presidential debate. This period represents a highly salient and politically charged moment, and the observed PW patterns may reflect dynamics amplified during election cycles rather than stable features of everyday discussion. Platform architecture and user demographics also vary, so comparative analyses across platforms, cultural contexts, and routine versus high-salience periods are needed.

Second, our reliance on automated tools for toxicity detection and target classification introduces bias. Such models often miss nuanced forms of incivility, leading to underestimation or mischaracterization. Combining large-scale computational methods with human annotation or qualitative analysis would improve accuracy and reveal more subtle rhetorical strategies.

Third, our definition of partisan warriors is behavioral--crossing ideological lines to post uncivil comments directed at identifiable partisan targets--and infers but does not explain intent. Some users may see themselves as debating, others as trolling or seeking attention. Surveys or interviews with commenters could clarify these motivations and provide a richer theoretical account.

Fourth, our findings are correlational and capture a single political moment. While toxic video titles are associated with more partisan warriors in conservative spaces, causal direction remains unclear. Longitudinal research tracking the same channels and commenters across election and non-election periods would help assess the temporal stability of PW behavior and disentangle short-term mobilization effects from enduring interaction patterns.

Fifth, future work could extend our analysis by modeling the impact of platform recommendation and visibility dynamics on commenting behavior. For example, integrating data on recommendation pathways, ranking positions, or personalized exposure signals may clarify how algorithmic curation interacts with content toxicity and partisan orientation. Such approaches would complement our content-focused analysis by accounting for exposure in shaping engagement patterns.

Finally, our dataset reflects only comments available through YouTube’s API at the time of collection. Because YouTube routinely removes content that violates its guidelines, the most extreme comments were likely already deleted. As a result, our analysis is necessarily conservative: we capture patterns among comments that remained visible rather than the full spectrum of hostility. Future work could address this by combining API-based datasets with real-time or pre-moderation collection strategies, enabling a more complete assessment of how toxicity unfolds before moderation.

\section{Acknowledgements}
We are grateful to Jason Greenfield (Computer Science PhD student, Princeton University) for his significant insights and helpful comments, which substantially improved the quality of this manuscript.

\bibliography{aaai2026}
\section{Paper Checklist}

\begin{enumerate}
\item For most authors...
\begin{enumerate}
    \item  Would answering this research question advance science without violating social contracts, such as violating privacy norms, perpetuating unfair profiling, exacerbating the socio-economic divide, or implying disrespect to societies or cultures?

   \answerYes{Yes, and we took several steps to address privacy concerns: (1) we analyzed only publicly posted comments accessible via YouTube's official API, (2) we focused on aggregate patterns rather than individual users, (3) we removed direct user identifiers in our analysis, and (4) our research aims to understand democratic discourse patterns rather than profile individuals.} 
  \item Do your main claims in the abstract and introduction accurately reflect the paper's contributions and scope?
    \answerYes{Yes}
   \item Do you clarify how the proposed methodological approach is appropriate for the claims made? 
    \answerYes{Yes, see Introduction}
   \item Do you clarify what are possible artifacts in the data used, given population-specific distributions?
    \answerYes{Yes}
  \item Did you describe the limitations of your work?
    \answerYes{Yes, see Limitations and Future Work}
  \item Did you discuss any potential negative societal impacts of your work?
    \answerYes{Yes, see Ethical Statements}
      \item Did you discuss any potential misuse of your work?
    \answerYes{Yes, see Ethical Statements}
    \item Did you describe steps taken to prevent or mitigate potential negative outcomes of the research, such as data and model documentation, data anonymization, responsible release, access control, and the reproducibility of findings?
    \answerYes{Yes, see Ethical Statements}
  \item Have you read the ethics review guidelines and ensured that your paper conforms to them?
    \answerYes{Yes}
\end{enumerate}

\item Additionally, if your study involves hypotheses testing...
\begin{enumerate}
  \item Did you clearly state the assumptions underlying all theoretical results?
    \answerNA{NA}
  \item Have you provided justifications for all theoretical results?
    \answerNA{NA}
  \item Did you discuss competing hypotheses or theories that might challenge or complement your theoretical results?
    \answerNA{NA}
  \item Have you considered alternative mechanisms or explanations that might account for the same outcomes observed in your study?
    \answerYes{Yes}
  \item Did you address potential biases or limitations in your theoretical framework?
    \answerYes{Yes, see Limitations and Future Work}
  \item Have you related your theoretical results to the existing literature in social science?
    \answerYes{Yes, see Related Work}
  \item Did you discuss the implications of your theoretical results for policy, practice, or further research in the social science domain?
    \answerYes{Yes, see Conclusion and Discussion}
\end{enumerate}

\item Additionally, if you are including theoretical proofs...
\begin{enumerate}
  \item Did you state the full set of assumptions of all theoretical results?
    \answerNA{NA}
	\item Did you include complete proofs of all theoretical results?
    \answerNA{NA}
\end{enumerate}

\item Additionally, if you ran machine learning experiments...
\begin{enumerate}
  \item Did you include the code, data, and instructions needed to reproduce the main experimental results (either in the supplemental material or as a URL)?
    \answerNA{NA}
  \item Did you specify all the training details (e.g., data splits, hyperparameters, how they were chosen)?
    \answerNA{NA}
     \item Did you report error bars (e.g., with respect to the random seed after running experiments multiple times)?
    \answerNA{NA}
	\item Did you include the total amount of compute and the type of resources used (e.g., type of GPUs, internal cluster, or cloud provider)?
    \answerNA{NA}
     \item Do you justify how the proposed evaluation is sufficient and appropriate to the claims made? 
    \answerNA{NA}
     \item Do you discuss what is ``the cost`` of misclassification and fault (in)tolerance?
    \answerNA{NA}
  
\end{enumerate}

\item Additionally, if you are using existing assets (e.g., code, data, models) or curating/releasing new assets, \textbf{without compromising anonymity}...
\begin{enumerate}
  \item If your work uses existing assets, did you cite the creators?
    \answerYes{Yes, see Methods}
  \item Did you mention the license of the assets?
    \answerNA{NA}
  \item Did you include any new assets in the supplemental material or as a URL?
     \answerNo{No}
  \item Did you discuss whether and how consent was obtained from people whose data you're using/curating?
    \answerYes{Yes. Data used was curated from publicly available sources.}
  \item Did you discuss whether the data you are using/curating contains personally identifiable information or offensive content?
     \answerYes{Yes, and we conform to the YouTube policies. See Ethical Statements.}

\item If you are curating or releasing new datasets, did you discuss how you intend to make your datasets FAIR?
\answerNA{NA}
\item If you are curating or releasing new datasets, did you create a Datasheet for the Dataset? 
\answerNA{NA}
\end{enumerate}

\item Additionally, if you used crowdsourcing or conducted research with human subjects, \textbf{without compromising anonymity}...
\begin{enumerate}
  \item Did you include the full text of instructions given to participants and screenshots?
    \answerNA{NA}
  \item Did you describe any potential participant risks, with mentions of Institutional Review Board (IRB) approvals?
    \answerNA{NA}
  \item Did you include the estimated hourly wage paid to participants and the total amount spent on participant compensation?
    \answerNA{NA}
   \item Did you discuss how data is stored, shared, and deidentified?
   \answerNA{NA}
\end{enumerate}

\end{enumerate}

\appendix
\section{Appendix}

\subsection{Robustness Check: User Categorization using Continuous Metric $\tau$}

\begin{figure}[h]
    \centering
    \includegraphics[width=\linewidth]{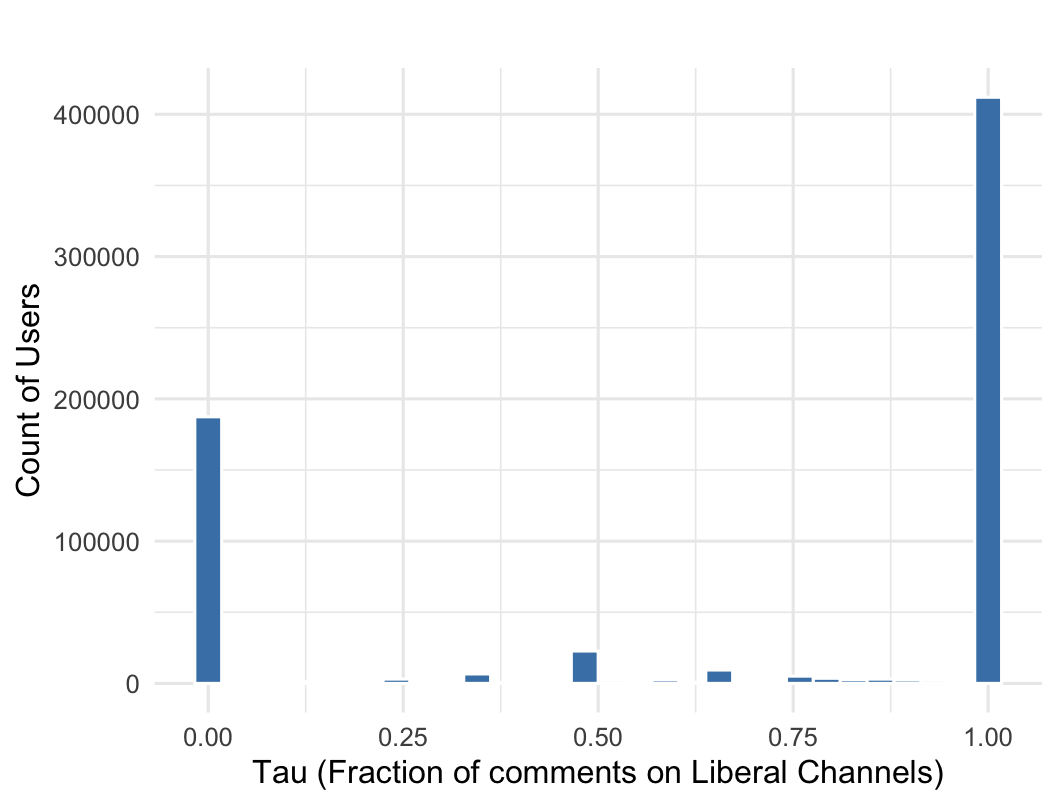} 
    \caption{Distribution of $\tau$ for All Users. The overall population exhibits a bimodal distribution, justifying categorical separation.}
    \label{fig:tau_all_distribution}
\end{figure}
\begin{figure}[h]
    \centering
    \includegraphics[width=\linewidth]{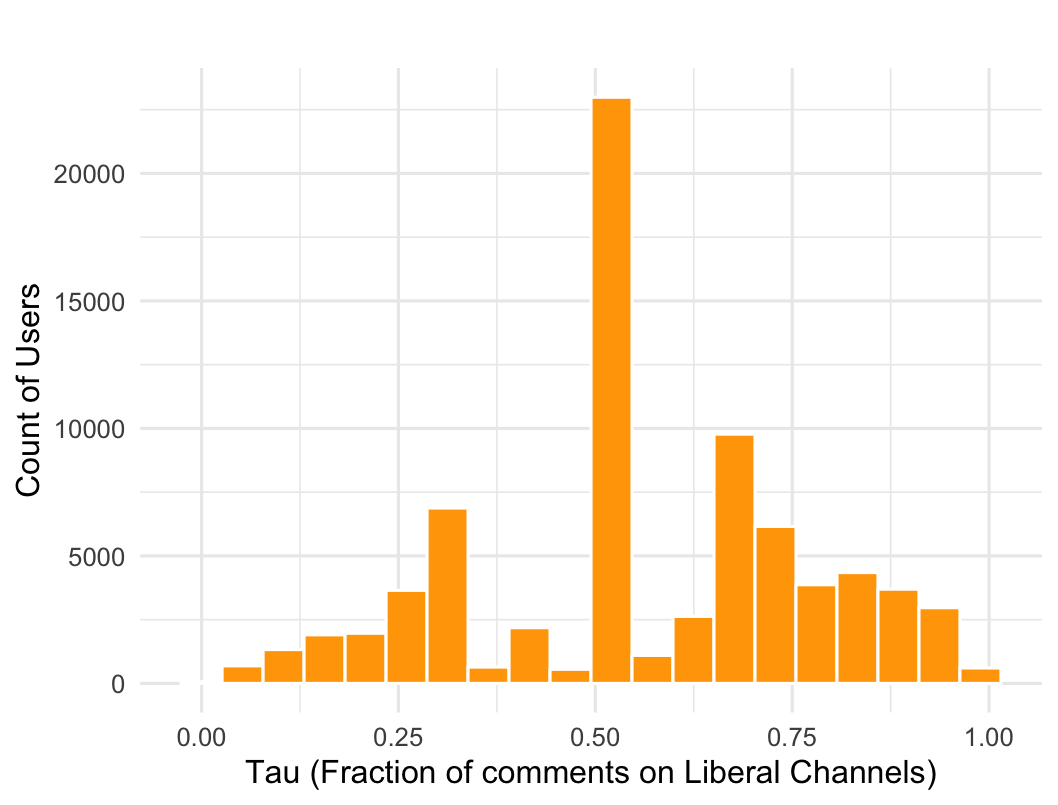}
    \caption{Distribution of $\tau$ among Cross-Commenters. The distribution peaks around $\tau = 0.5$, indicating balanced engagement across ideological contexts.}
    \label{fig:tau_cross_distribution}
\end{figure}

To assess the robustness of our categorical user definitions (L-Only, C-Only, and Cross-Commenters), we additionally computed a continuous engagement metric, $\tau$, defined as the fraction of a user's comments posted on liberal-leaning channels. Under this definition, $\tau = 1$ corresponds to exclusive participation on liberal channels, whereas $\tau = 0$ indicates exclusive participation on conservative channels.

Figure~\ref{fig:tau_all_distribution} presents the distribution of $\tau$ for the full commenter population ($N = 12{,}884$). The distribution is strongly bimodal, with the majority of commenters concentrated at the extremes ($\tau = 0$ or $\tau = 1$). This pattern indicates that most commenters engage almost exclusively within a single ideological context, supporting the use of categorical classifications to capture dominant participation patterns.

Figure~\ref{fig:tau_cross_distribution} shows the distribution of $\tau$ among commenters classified as Cross-Commenters. Rather than clustering near either extreme, this group exhibits a broad distribution of $\tau$ values with a clear mode around $\tau = 0.5$. This indicates that Cross-Commenters tend to engage substantially with both liberal- and conservative-leaning channels, rather than predominantly favoring one side with minimal cross-ideological activity.

Taken together, these results suggest that ideological engagement on YouTube is best characterized by discrete participation regimes rather than a smooth continuum. While $\tau$ provides a useful descriptive measure, the underlying user behavior is highly polarized, and the categorical distinctions employed in our analysis reflect meaningful and empirically grounded differences in commenting patterns.

\subsection{Characterizing `Unclear' Commenters}
\label{sec:appendix_unclear}

To address the nature of users classified as \textit{Unclear} ($-0.2 < \text{stance\_score} < 0.2$), we examined their prevalence and behavioral patterns. As shown in Table \ref{tab:stance_validity}, these users constitute a negligible fraction of the active user population across all groups.

Qualitative inspection reveals that these users are typically not ``trolls'' attacking both sides, but rather commenters whose hostility falls outside the specific Trump-Harris binary defined in our taxonomy. Their comments generally fall into two categories:

\begin{enumerate}
    \item \textbf{Ambiguous References:} Comments relying on pronouns without sufficient context for the model to resolve a specific target (e.g., \textit{``This lady sucks,''} \textit{``He is an idiot''}). While likely partisan in context, they lack the explicit naming required for our conservative coding strategy.
    \item \textbf{Peripheral Targets:} Attacks directed at entities other than the main presidential candidates, such as media organizations or other political figures (e.g., \textit{``CNN is pathetic... Sick of the lies,''} \textit{``ABC is so biased and disgusting,''} \textit{``Nikki [Haley] you are a two-faced snake!''}).
\end{enumerate}

Thus, the classification of \textit{Unclear} largely reflects the strict operationalization of our target taxonomy (focusing on the Trump vs. Harris rivalry) rather than a lack of partisan sentiment among these users.

\subsection{Prompt example for detecting attack target from comments}

\begin{spverbatim}
You are an assistant who identifies the target of uncivil political YouTube comments and classifies the target.

Your job is to:
- Identify only the ATTACK target(s) — entities mentioned or implied in the comment with uncivil language. Uncivil language is a feature of discussion that conveys an unnecessarily disrespectful tone toward the discussion forum, its participants, or its topics.
- For each target, classify it into one or more of the following categories:
0. Unclear - The referent is absent, or cannot be confidently resolved, even with metadata.
1. Supporters of Donald Trump
2. Supporters of Kamala Harris
3. Kamala Harris (as an individual)
4. Donald Trump (as an individual)
5. Republicans
6. Democrats
7. Other political actors
8. Conservatives in general
9. Liberals in general
10. Individual or group(s) with a particular religion
11. Individual or group(s) with particular gender identities
12. Individual or group(s) with particular race/ethnicity
13. Immigrants
14. Debate moderators
15. Mainstream media or journalist(s) associated with mainstream media (Specify the media)
16. Government Bodies (White House)
17. Joe Biden
18. Others - Other specified targets that cannot be classified from 1 - 17. 

- For vague references (e.g., "he", "she", "they"), ONLY resolve them using the video metadata when there is strong evidence for a specific resolution. If there is any ambiguity, classify the target as "Unclear" (category 0).
- ONLY identify targets that are being attacked with uncivil language, not just mentioned or criticized in civil terms.
...

\end{spverbatim}

\subsection{Methodological Pipeline}

\begin{figure}[H]
    \centering
    \includegraphics[width=0.9\columnwidth]{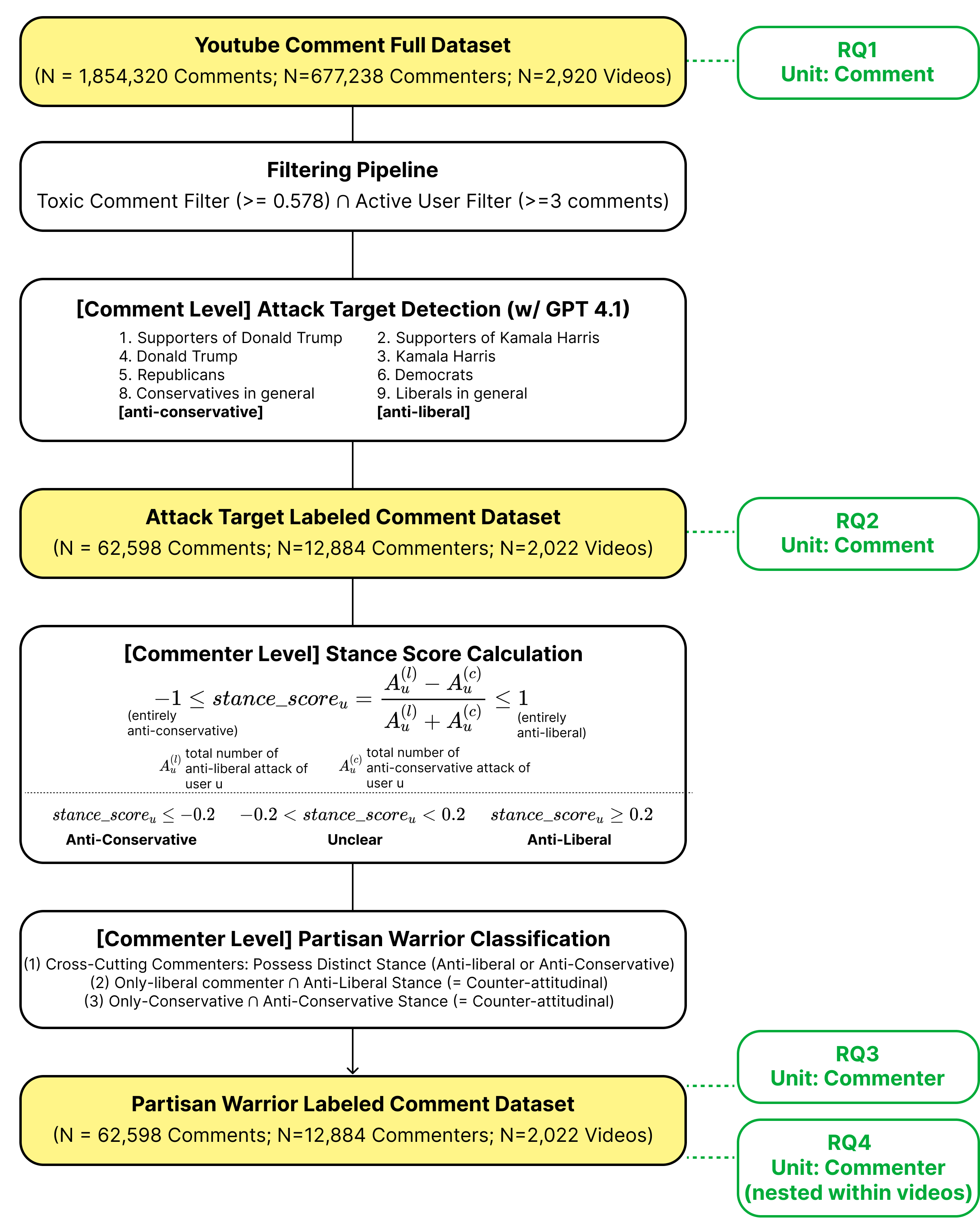}
    \caption{Methodological pipeline showing the complete process from YouTube comment collection (N=1,854,320) to partisan warrior classification.}
    \label{fig:pipeline}
\end{figure}

\subsection{Supplementary Analyses}

\begin{table*}[h]
\centering
\begin{tabular}{cccccccccc}
\toprule
\multirow{2}{*}{Threshold $T$} & \multirow{2}{*}{N} &
\multicolumn{3}{c}{Stance Distribution (\%)} &
\multicolumn{4}{c}{PW Rate (\%)} &
\multirow{2}{*}{$\chi^2$ (dof=2)} \\
\cmidrule(lr){3-5} \cmidrule(lr){6-9}
 & & Anti-Lib. & Anti-Con. & Unclear & Overall & OnlyL & OnlyR & Both & \\
\midrule
0.2 & 12{,}884 & 20.4 & 73.6 & 6.0 & 48.0 & 6.6 & 27.3 & 93.6 & 9222.1$^{***}$ \\
0.3 & 12{,}884 & 20.4 & 73.5 & 6.2 & 47.9 & 6.5 & 27.2 & 93.4 & 9198.9$^{***}$ \\
0.4 & 12{,}884 & 20.1 & 72.9 & 7.1 & 47.4 & 6.4 & 27.0 & 92.4 & 9016.8$^{***}$ \\
0.5 & 12{,}884 & 20.1 & 72.8 & 7.1 & 47.4 & 6.4 & 27.0 & 92.3 & 9009.9$^{***}$ \\
0.6 & 12{,}884 & 19.9 & 72.1 & 8.0 & 46.8 & 6.3 & 26.4 & 91.2 & 8820.1$^{***}$ \\
0.7 & 12{,}884 & 19.7 & 71.3 & 9.0 & 46.1 & 6.2 & 26.2 & 90.0 & 8602.4$^{***}$ \\
0.8 & 12{,}884 & 19.6 & 70.9 & 9.6 & 45.8 & 6.1 & 25.7 & 89.4 & 8496.3$^{***}$ \\
\bottomrule
\end{tabular}

\begin{flushleft}
\footnotesize
Notes: Anti-Lib. = anti-liberal, Anti-Con. = anti-conservative.
PW Rate indicates the proportion of commenters classified as partisan warriors within each commenter type.
$\chi^2$ values are from chi-squared tests of independence between commenter type and PW status.
$^{***}p < .001$.
\end{flushleft}

\caption{Robustness check of stance thresholding. Distribution of stance labels and partisan warrior (PW) rates across thresholds. All proportions are computed from the same set of 12,884 commenters.}
\label{tab:threshold}
\end{table*}

\begin{figure*}[h]
\centering
\includegraphics[width=0.95\textwidth]{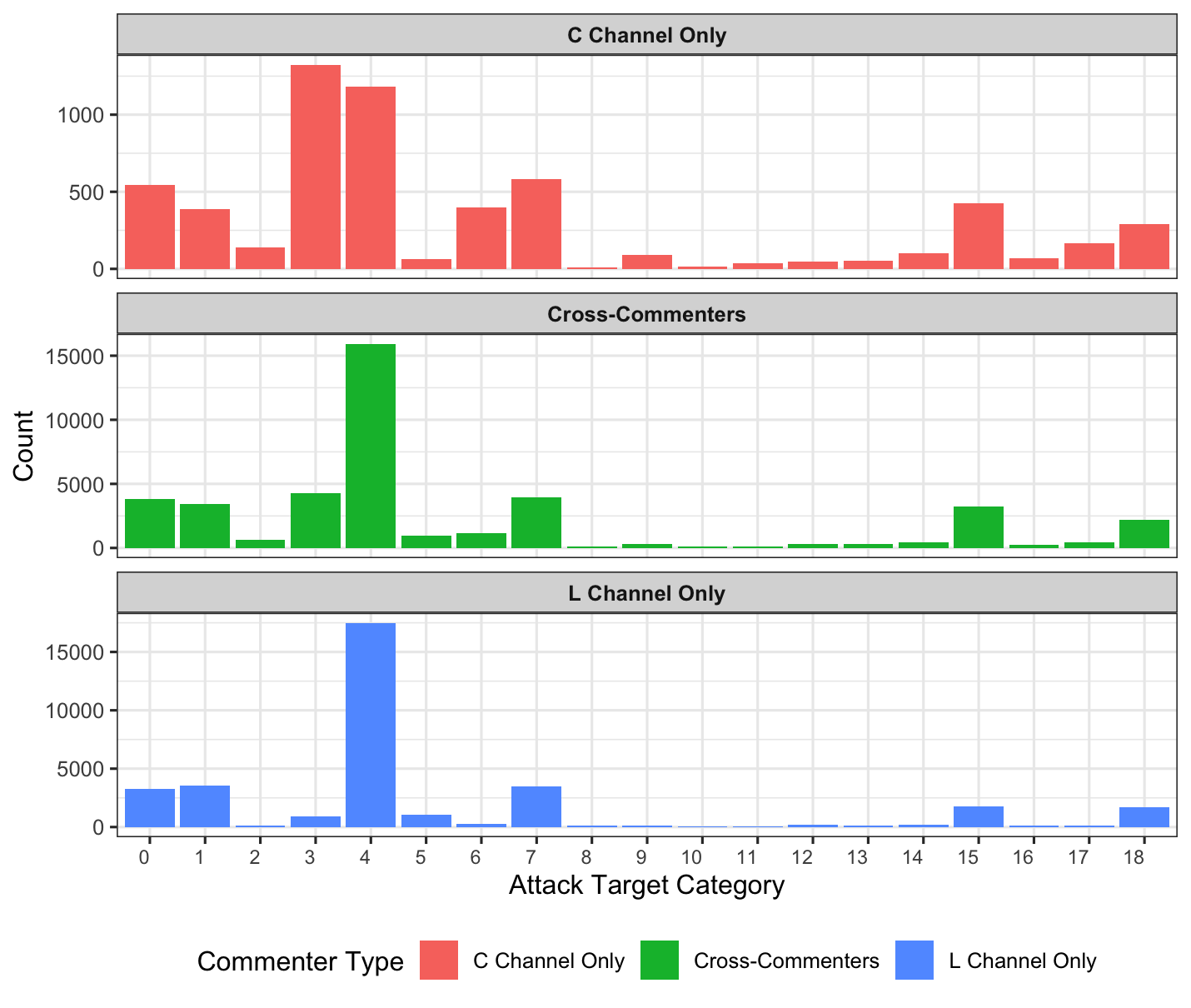} 
\caption{Distribution of total attack targets across commenter types, defined by the channel on which comments were posted (C Channel Only, L Channel Only, Cross-Commenters). Bars represent the total number of uncivil attacks directed at each target category.}
\label{fig:toxicity_all}
\end{figure*}

\end{document}